**Epidemiologic analyses with error-prone exposures: Review of current practice and recommendations**

Pamela A. Shaw[1], Veronika Deffner[2], Ruth H. Keogh[3], Janet A. Tooze[4], Kevin W. Dodd[5], Helmut Küchenhoff[2], Victor Kipnis[5], Laurence S. Freedman[6] on behalf of Measurement Error and Misclassification topic group (TG4) of the STRATOS Initiative

Affiliations:

1. Department of Biostatistics, Epidemiology and Informatics, University of Pennsylvania Perelman School of Medicine, Philadelphia, PA 19104, USA.

2. Statistical Consulting Unit StaBLab, Department of Statistics, Ludwig-Maximilians-Universität, Munich, Germany

3. Department of Medical Statistics, London School of Hygiene and Tropical Medicine, London, UK.

4. Department of Biostatistical Sciences, Wake Forest School of Medicine, Medical Center Blvd, Winston-Salem, NC, 27157, USA.

5. Biometry Research Group, Division of Cancer Prevention, National Cancer Institute, 9609 Medical Center Drive, Bethesda, Maryland 20892, USA.

6. Information Management Services, Inc., 6110 Executive Boulevard, Rockville MD 20852, USA and Biostatistics Unit, Gertner Institute for Epidemiology and Health Policy Research, Sheba Medical Center, Tel Hashomer 52621, Israel.

Corresponding author: Pamela Shaw, 606 Blockley Hall, Department of Biostatistics, Epidemiology, and Informatics; University of Pennsylvania Perelman School of Medicine; Philadelphia, PA 19104, USA; email: shawp@upenn.edu; Tel: 215-898-9599.

Conflicts of interest: None.

Sources of financial support: RHK is supported by a Medical Research Council Fellowship (MR/M014827/1). JAT is supported in part by NCI Cancer Center Support Gran*t* P30 CA012197*.*

**Abstract:**

Background: Variables in epidemiological observational studies are commonly subject to measurement error and misclassification, but the impact of such errors is frequently not appreciated or ignored. As part of the STRengthening Analytical Thinking for Observational Studies (STRATOS) Initiative, a Task Group on measurement error and misclassification (TG4) seeks to describe the scope of this problem and the analysis methods currently in use to address measurement error.

Methods: TG4 conducted a literature survey of four types of research studies that are typically impacted by exposure measurement error: 1) dietary intake cohort studies, 2) dietary intake population surveys, 3) physical activity cohort studies, and 4) air pollution cohort studies. The survey was conducted to understand current practice for acknowledging and addressing measurement error.

Results: The survey revealed that while researchers were generally aware that measurement error affected their studies, very few adjusted their analysis for the error. Most articles provided incomplete discussion of the potential effects of measurement error on their results. Regression calibration was the most widely used method of adjustment.

Conclusions: Even in areas of epidemiology where measurement error is a known problem, the dominant current practice is to ignore errors in analyses. Methods to correct for measurement error are available but require additional data to inform the error structure. There is a great need to incorporate such data collection within study designs and improve the analytical approach. Increased efforts by investigators, editors and reviewers are also needed to improve presentation of research when data are subject to error.

**Introduction**

Measurement error is a challenge in many settings in epidemiology. Exposures such as dietary intakes, environmental contaminants, and physical activity are difficult to measure because patterns of exposure are complex and because accurate (unbiased) and precise (with minimal variability) ways to measure many such exposures of interest are either not available or are too impractical to use in a large study. Practical measures for these exposures will contain random deviations from a target exposure, such as a short-term mean dietary intake, due to biological variability or assay error, and also potentially systematic bias, for example from inaccurate self-reported exposures. Here, we refer to both of these kinds of deviations from the target exposure as *measurement error*; with random error defined as mean zero, independent error and systematic error defined as covariate dependent errors. In some cases, through intensive monitoring and/or better instruments, more accurate and precise measurements that assess the measurement error structure of a study instrument are available in a subset of subjects or from an independent validation study. These reference, or gold standard, measures can inform a method that corrects for the instrument error[1-6] or a quantitative bias analysis[7-11]. It is well-established in the statistical and epidemiological literature that if the measurement error in an exposure variable is ignored, analyses can be subject to biased estimation and incorrect inference[1-11].

Analysis techniques to address exposure measurement error have been the focus of methodologic research in statistics and epidemiology for several decades. These efforts have produced many methodological advances in both analysis techniques and study designs. A number of these methods have been summarized in statistical[1-4] and epidemiological texts[5,6]. Several review papers on how to use existing methods or that compare methods in specific settings have also appeared in the literature[12-16]. Additionally, there have been many articles appearing in epidemiologic journals advocating that quantitative bias analyses be provided for any analysis that involves error-prone exposure measures[7-11]. Despite these efforts a surprising number of articles have been published in the biomedical literature with no adjustment in the data analysis and little to no discussion of how measurement error potentially impacted the study results[17]. This has been true even in research areas, such as nutritional epidemiology, where there is a well-established literature in topic matter journals, instrument-specific software[18] and webinars to make these methods more accessible[19].

The international STRengthening Analytical Thinking for Observational Studies (STRATOS) Initiative is a large collaboration of experts in many different areas of biostatistical research that was formed in response to an observation that many methodological advances in statistics are not put to practice and that the design and analysis of observational studies commonly exhibit serious weaknesses[20]. The objective of STRATOS is to provide accessible and accurate guidance documents for relevant topics in the design and analysis of observational studies. The STRATOS Initiative to date has formed working groups in 9 topics: missing data, selection of variables and functional forms in multivariate analysis, initial data analysis, measurement error and misclassification, study design, evaluating diagnostic tests and prediction modeling, causal inference, survival analysis, and high dimensional data.

In this article, we present the results of a literature survey done by the STRATOS Measurement Error and Misclassification Topic Group (TG4) to assess current practice for handling measurement error in the biomedical literature. We performed literature surveys in four areas of epidemiology where measurement error is a well-known concern: (1) dietary intake cohort studies, (2) dietary population surveys, (3) physical activity cohort studies, and (4) air pollution cohort studies. In the cohort studies, we were specifically interested in analyses of the association between an error prone exposure and outcome, and in the dietary population surveys we were specifically interested in analyses used to estimate the distribution of intake of a dietary component (a nutrient or food). We focused on mis-measured exposures (and not outcomes), because this has been the focus of the majority of statistical methods work, and software, to address measurement error. We present the results of these literature surveys. We describe whether investigators are using appropriate statistical methods to adjust for or assess the potential effects of measurement error on study results, and to what degree authors do or do not discuss the impact of measurement error on their results. We also describe which methods are used by those who address measurement error in their analysis. We conclude the article with recommendations for how to overcome the short-comings in current practice for statistical analyses, and consequently in the resulting scientific conclusions, in fields where measurement error remains a challenge.

**Methods**

*Overview*

STRATOS TG4 conducted a literature survey of four research areas: 1) dietary intake cohort studies, 2) dietary intake population surveys, 3) physical activity cohort studies, and 4) air pollution cohort studies. For the three cohort study literature surveys, articles for review were identified by two types of search: A) a search with general search terms related to the topic area and B) a similar search with additional required terms related to measurement error or misclassification. The purpose of search A was to conduct a general review of the topic areas to understand the current practice for how error-prone exposures are handled. Search B, performed only for the cohort studies, was done in expectation that few articles from search A would have a measurement error adjusted analysis. The purpose of search B was to identify articles that in some way did address measurement error or misclassification in the analysis, in order to be able to summarize which methods are currently in use. For dietary cohort studies, using a method to adjust for the mis-measured exposure was required to be eligible for Search B; due to a lack of authors applying measurement error methods, this was not required for the other topic areas. For the dietary intake survey literature, only search A was performed because, while issues of variability around usual intake are appreciated in this setting, the terms misclassification or measurement error are typically not used.

A general protocol that specified the questions to be addressed in the literature review across the topic areas was developed in advance [Supplemental Materials]. Questions included what was the statistic of primary interest, whether the article mentioned measurement error in the exposure variable as being a potential problem, whether a reliability (repeated measures) or calibration (comparison with a reference instrument) sub-study was included, and whether any methods were used to address measurement error in the data analysis. Separate literature surveys were performed by topic area and tailored to increase relevance for each specific setting. For the cohort studies, the error prone measure of interest was required to be an exposure and not an outcome in the presented analyses. For the dietary intake and physical activity cohort surveys, questions regarding whether exposures were analyzed as categorical or continuous variables were included. In those surveys, we also collected information on whether multiple exposures subject to error were included in the regression analysis, focusing on the common examples of 1) physical activity and dietary intake exposures and 2) dietary intake and smoking. Data extraction instruments were reviewed by the TG4 working group before initiation of the literature search. The literature search was performed in 3 stages: 1) identify appropriate research articles using PubMed/Web of Science, 2) review titles and abstracts to select articles in scope, and 3) detailed review of selected articles for data abstraction, making further exclusions if necessary. Meta-analyses, review articles, interventional studies, and retrospective case-control studies were excluded. Further details by topic area are provided below.

Reviews of Search A and Search B articles were done by one or more primary reviewers, generally one reviewer per article. For purposes of quality control (QC) a 20% subsample, stratified by reviewer and search type, was randomly chosen for review by a second reviewer. PRISMA flow diagrams are provided in the Supplemental Tables and Figures[21].

*Dietary Intake Cohort Study*

For search A, a literature search with general search terms including "nutritional intake" or "dietary intake" and "cohort" identified tens of thousands of articles. Search A was restricted to the most recent 12 months and a few disease areas for which diet is a well-known exposure of interest, namely one of "cancer", "cardiovascular disease", and "diabetes". Search A also included the term "risk" to try to separate studies of associations between dietary exposures and health outcomes from the many articles considering adequacy of intake for a certain population. The ePub date June 2, 2014-June 2, 2015 captured 51 articles. Search terms are provided in Table 1.

Search B required the terms "measurement error" or "misclassification" and was expanded to 3 different searches and the previous 15 years in order to identify a target of 30 qualifying articles (criteria in Table 1). Ultimately 31 search B articles with ePub dates January 1, 2001- July 15 2015 were selected for full review. Supplemental Figures 1 and 2 provide the number of articles initially returned by each stage of the search for search A and B, respectively. The randomly selected QC subsample included 10 search A and 6 search B articles, split equally by two primary reviewers (RHK, PAS) and reviewed independently by a third reviewer (VK).

*Dietary Intake Survey*

Articles were identified in one query in Pubmed with date range: 01/01/2012 – 05/31/2015 and search terms that included a variant of the word "nutrition" and terms related to typical dietary intake survey instruments (Table 1). The query returned 2801 articles. Title and abstract review was performed by a single primary reviewer (KWD) on the most recent 717 identified. The review was restricted to surveys whose aim was to describe a population of some geographic region. The first 67 articles identified as within scope were given detailed review; all 67 were confirmed eligible and data were extracted. Supplemental Figure 3 provides the PRISMA flow diagram. A QC sample of 13 papers was randomly selected and reviewed by a second, independent reviewer (LSF).

*Physical Activity Cohort Study*

For search A, a query in PubMed was performed with date range 07/01/2012 – 06/30/2015 and required terms that aimed to narrow the search to prospective cohort studies with physical activity exposures (see Table 1). For Search B, terms relating to measurement error and misclassification were added to the query required terms (Table 1). There were 8760 articles returned from search A and 610 from search B. We selected a random subset of search A articles equal to the number of search B articles identified. After abstract review, there were 50 articles from search A and 87 from search B determined to be eligible for data abstraction. Fifty articles from each search (a random subset for search B) were abstracted for review. Upon detailed review, there were 30 articles abstracted for search A and 39 articles for search B. Supplemental Figures 4 and 5 provide the PRISMA flow diagrams for searches A and B, respectively. Reviews were done by a single primary reviewer (JAT) and a QC sample of 10 randomly selected search A and 10 search B articles were reviewed by an independent reviewer (LSF).

*Air Pollution Cohort*

For search A, a query in Web of Science was performed with date range 01/01/2012 – 12/31/2014 and the search terms shown in Table 1. The search was restricted to prospective cohort studies with health outcomes (e.g. cardiovascular disease, cancer mortality, hospital admissions and respiratory disease) that are often subjects of research on the impact of air pollution on human health. For search B the term "health" was added to the measurement error terms. In addition, a very general search with the terms *"measurement error" and "air pollution"* was conducted.

Search A returned 4682 articles; search B returned 386 articles. After title/abstract review there were 451 eligible articles for Search A and 32 for search B. For search A, randomly selected articles were read in detail (MA,VD,AH,HK,TM) and data were extracted until 50 eligible articles were found. For search B, all 32 articles were reviewed (VD) and 25 found eligible upon detailed review. Supplemental Figures 6 and 7 provide the PRISMA flow diagrams for search A and B, respectively. A randomly selected QC sample of 10 papers for search A and 5 papers for search B had data extracted by an independent reviewer (NH,MA).

**Results**

We describe the results of each literature review separately. Tables 2 and 3 summarize the main results for search A and B, respectively.

*Dietary Intake Cohort Study*

In the general review (search A), 46/51 (90%) of articles analyzed the nutritional intake exposures with no adjustment for measurement error in the analysis. Of the 5 that performed some adjustment, 2 (4%) used regression calibration[1,22] and 3 (6%) used a cumulative average of multiple assessments over time. Most authors (48, 94%) did acknowledge in some way that their exposure was prone to measurement error. Only 31 (61%) mentioned that their reported association corresponding to the error-prone exposure was subject to bias, whereas the remaining articles either made no or a vague reference to errors in the exposure measurement, such as that their instrument had been validated and had an acceptable reliability coefficient, which may or may not have been provided. When measurement error was mentioned, generally an incomplete or false claim was made about its impact on the presented analysis, such as the effect estimates were subject to attenuation bias only or their study was not subject to bias because it was prospective or because their instrument had good reliability. None described adjusting the study design to accommodate the error in the study measurement, i.e. increasing sample size to offset loss of power. Nearly all articles (50, 98%) categorized their error-prone exposure and 27 (53%) did so exclusively.

For search B, 4 of 31 articles upon detailed review were found not to have addressed exposure error in the analysis and were excluded; 3 reported no method and one reported using only an energy-adjustment method. This latter article was discarded from the eligible search B articles as energy-adjustment methods are generally not considered a formal method to address random or systematic error in the exposure. Thus, 27 articles were included in the reported search B analysis. All but 1 of the articles (96%) used regression calibration to address the error; one of these articles additionally used SIMEX.[1,23,24] The remaining article considered both an average of multiple measures and a latent variable technique as alternative analysis approaches. Amongst the 26 articles that used regression calibration, 6 were informed by a reliability substudy (repeated measures of the same instrument), 14 calibrated to a different self-report instrument or food record (FR), 4 calibrated to an objective recovery biomarker, 1 performed calibration separately by 3 different instruments (biomarker, FFQ, FR) and one did not report the calibration instrument. Twenty-four (89%) search B articles reported using a method that adjusted the standard error estimation for the extra uncertainty induced by the measurement error adjustment. None discussed considering the error in the study design.

For search B articles, there were no explicitly false claims about error, but 44% had an incomplete discussion. Incomplete discussions included failing to mention the limitations of using a reference instrument with errors correlated with those of the main study instrument and failing to acknowledge that calibrating only for within-person variability may have been inadequate due to systematic errors in the main study instrument (such as the FFQ).

Several articles had more than one error-prone exposure in their association analysis. Seventy-six percent of search A and 81% of search B articles reported having considered self-reported physical activity as an exposure in a multivariate regression, in addition to the dietary exposure; none of the search A and 2 (7%) of the search B articles explicitly reported adjusting their analysis for the errors in both the physical activity and nutritional intake exposures. There were 92% (89%) of the search A (B) articles that additionally adjusted for self-reported smoking, generally also considered to be error-prone. None adjusted for errors in the smoking variable.

*Dietary Intake Survey*

The selected dietary intake survey papers provided analyses addressing one or more of the following themes: 1) investigating differences in measures of central tendency (mean/median) of intake or fraction consuming specific dietary components 2) ranking food/food group contributors to overall diet, and 3) examining the distribution of intake, in particular: percentiles other than the median, fractions consuming less or more than specified limits (such as a recommended daily intake), and quantities such as a dietary pattern score that are derived from the two former types of statistics.

Most (53/67, 79%) of the articles assessed dietary intake using one or more 24-hour recalls. Of the remaining 14 articles, 3 analyzed diaries/food records, and 11 analyzed FFQs. Of these 14, all but one (13/14, 93%) recognized that the self-report instrument used could be subject to underreporting or bias, but generally (10/13, 77%) such recognition was presented only in the Discussion section of the paper as a limitation. Of the 53 papers concerned with 24-hour recalls, 32 (60%) presented analyses of a single recall (even if multiple recalls were available on some respondents), and 21 (40%) presented analysis including multiple recalls on at least some respondents. Both 1-recall and multi-recall papers were likely (69%/86%) to note that 24-hour recalls could be subject to underreporting/systematic bias due to measurement error. Complex modeling methods are available to estimate the full distribution of usual intake from 24-hour recall data, considering features such as within-subject variability and episodic consumption.[25-30] These methods can also be used to estimate fractions with usual intake below fixed cut-points, or distributions of scores ostensibly computed on usual intake. However, these methods generally cannot be applied when only 1 24-hour recall per respondent is available. Nevertheless, 10/32 (31%) did present statistics other than the mean, which were subject to bias. By comparison, 15 of the 21 papers with multiple recalls (71%) also presented these types of statistics (though they did not necessarily utilize the complex methods). Half (16/32, 50%) of the 1-recall papers made no mention of within-person variation or usual intake, compared to only 5/21 (24%) of the multi-recall papers. When the 1-recall papers mentioned within-person variation or usual intake, it was often in the context of justifying their analysis on the basis that a single recall can be used to estimate the mean of the usual intake distribution under the classical (independent and unbiased) measurement error assumption. Overall, only 19/67 (28%) used a method to adjust their analyses for measurement error (Tables 2,3).

*Physical Activity Cohort Study*

The majority of studies examined in searches A (24, 80%) and B (37, 95%) were prospective cohort studies. A number of different constructs of physical activity were measured in both searches, with the most common being minutes of moderate or vigorous activity (Search A: N=11, 37%; Search B: N=9, 23%), sedentary activity (A: N=6, 20%; B: N=5, 13%), and adherence to guidelines (A: N=3, 10%; B: N=7, 18%). Other constructs included: metabolic equivalents (MET) minutes, activity energy expenditure, total energy expenditure, and various scale-based or ad hoc measures of summarizing activity. Most authors only used subjective measures of activity (A: N=25, 83%; B: N=33, 87%); a small number (A: N=4, 13%; B: N=4, 11%) used only objective measures, such as those from an accelerometer; and one from each search used both.

In Search A, 70% (N=21) categorized the physical activity variable; in Search B, 79% (N=31) did.

None of the physical activity articles identified in the general search (Search A) analyzed physical activity with adjustment for measurement error. Over half of the papers in Search A (N=18/30) mentioned measurement error or misclassification as a limitation. Of these 18 papers, 13 mentioned bias due to self-report: of these, six mentioned attenuation, five mentioned that physical activity may be overreported, and four hypothesized that the error was most likely non-differential, i.e., not associated with the outcome of interest, and therefore likely to lead to attenuation, though in fact this is not necessarily the case. Only two papers mentioned designing the study to account for measurement error. None of the studies examined had a calibration substudy. In Search A, three studies (10%) included repeated measures, but only to assess change over time, not to address repeatability. In Search A, 37% (N=11) of the primary analysis regression models also adjusted for nutritional self-report exposures; 10 of these adjusted for alcohol exposure.

In Search B, five papers (13%) mentioned that measurement error was considered in some way in the study design; three of these (8%) had a calibration substudy, and two (5%) had an adjustment for measurement error. Both studies used a form of regression calibration. Overall, 27 papers (69%) mentioned measurement error as a limitation, and four papers specifically addressed reliability or validity of measures. Of these 27 papers, 21 mentioned bias due to self-report; seven mentioned measurement error may have attenuated the estimated relationship with the outcome; two mentioned that physical activity may be overreported; two mentioned power loss, and two hypothesized that the error was most likely non-differential. Interestingly, one paper acknowledged their use of questionnaires could result in error and residual confounding, with an unknown magnitude and direction of bias; and that categorization could reduce power, and introduce differential error.[31] Three studies (8%) used repeated measures to assess within-person variability. In Search B, 59% (N=23) of the primary analysis models also included adjustment for nutritional self-report data; all of them adjusted for alcohol.

*Air Pollution Cohort Study*

In the search A articles, an individual's exposure was predominantly measured at fixed-site monitoring stations (35, 70%) that recorded data on hourly or daily temporal resolution (30, 60%). Table 4 describes characteristics of the search A articles. In general, measurement error in air pollution cohort studies arises by temporal and/or spatial aggregation of the exposure data from a fixed monitor that is then applied to an individual, limited availability of exposure measurements with temporal and/or spatial variability, and error in the instruments themselves. Fewer than half the papers (23, 46%) mentioned measurement error as a potential problem, and those that did mention it did not describe its sources or its impact on the study results in detail. Four (8%) search A articles used a method to address measurement error in their analysis. One study used a type of regression calibration to account for systematic measurement error in the older of two instruments, but the method was not denoted as regression calibration. Three studies conducted sensitivity analyses. Two (4%) out of all the studies included a subsample in which a reference measure was available enabling examination of measurement error (one for regression calibration and one to measure reproducibility).

In the search B articles identified, the measurement error problem was predominantly only mentioned or discussed, but not formally analyzed. Only 5 (20%) studies applied methods to address the measurement error; all five studies used aggregated exposure values (fixed-site exposure measurements or estimated exposure values). A single study applied an instrumental variable approach to deal with potential measurement error; the authors of the other four studies performed sensitivity analyses to describe the robustness of their results under different assumptions concerning the amount of measurement error present.

**Discussion**

Dietary intake and physical activity exposures were typically collected using self-reported data, which can be subject to systematic errors dependent on the true value and classical (random) errors independent of the true value. The majority of air pollution studies used fixed-site monitoring stations, which are subject to complex error structures including systematic and classical errors, as above, but also Berkson error (i.e. error that is random and independent of the measured value, but arises due to aggregate rather than individual exposure being measured)[1]. Yet, for the majority of articles reviewed in the prospective cohort studies, there was an inadequate discussion of the impact of measurement error on study results. Several authors made no or only vague discussion of the measurement error, stating only that it was present in the exposure measurement but not being clear on its origin, size, structure or its potential impact on estimated associations. Consideration was also not given to the impact of dichotomizing a continuous error-prone variable, which can be unpredictable depending on the relationship between the unobserved exposure and outcome[32]. Amongst the authors that did discuss measurement error in the context of an association analysis, several incorrectly claimed that attenuation was the only possible direction of bias induced by error. Though for dietary intake exposures, it has been noted that for the predominant direction of bias has been observed to be attenuation, attenuation is not generally guaranteed in multivariable models – even with only random error [33]. Furthermore, adjusting for multiple error-prone exposures was common and the fact that this can influence the direction of the bias was generally not mentioned in any of the articles reviewed. Authors ignoring errors were also not adjusting their analysis for the level of uncertainty in the exposures, thus overstating the precision of the target association.

In nearly all the studies with multiple error-prone exposures, either no or only one was directly addressed. It was also clear that authors were not fully taking advantage of available information in their study to inform the structure of the measurement error. For example, in many cohort studies and the dietary intake surveys, several authors were not taking advantage of repeat measurements in settings where they could have been used to assess the impact of within-person variation on study estimates. Many also cited validation studies of the exposure instruments utilized in the study, which may have included simultaneous assessment with a reference instrument, and therefore may have been used for calibration.

There were a few topic-specific themes that also arose. In many dietary survey analyses, interest is in "usual intake", defined as long-term average daily intake of a dietary component. Based on the sample of dietary survey papers, researchers in that niche commonly use 24-hour recalls or food record methods of assessing dietary intake. Those who use the 24-hour recalls/records seem willing to make the working assumption that their instruments of choice are unbiased, though many concede in the Discussion section that the working assumption is probably violated. Even those applying the complex modeling methods to derive usual intake from the self-report instruments, in order to adjust their study estimates for error in the dietary intake measurements, generally attempt to adjust for only within-person variation and not systematic bias.

Our review of the physical activity literature found that multiple different constructs are used to describe physical activity.  Some constructs use time and not intensity, some focus only on leisure activities; some provide continuous measures, while most are categorized. The choice of construct is important with respect to measurement error, as different measures are subject to different types of bias; however regardless of construct, there was very little attention in the analysis or interpretation of results regarding the impact of measurement error.

The measurement error problem in air pollution cohort studies arises from a complex exposure and the error structure can vary considerably by study design. First, Berkson error is prevalent, arising in studies that commonly rely on aggregated fixed-site exposure measurements with low temporal resolution[1]. Only a few of these studies discuss the underlying assumptions for validity of estimated model parameters for exposures containing Berkson error. Comments about Berkson error are mostly confined to discussion of the spatial variability of the actual exposure; potential biases due to Berkson error in complex settings like air pollution studies are usually not discussed. Second, exposure measurements with classical measurement error occur in studies with personal exposure measurements; investigating or adjusting for the impact of measurement error on study results is not current practice in such studies, with none of the five identified articles that analyzed personal exposure measurements applying any adjustment for measurement error in their health outcome disease association studies.

**Conclusion**

The presented literature survey reveals that articles with inadequate treatment of exposure measurement error and misclassification in the analysis and discussion of their study results remain commonplace in the literature. We focused on covariates prone to mismeasurement, as this setting has been the dominant focus of existing methods to address measurement error; however, we expect similar problems exist in published analyses that also involve outcomes that are prone to error, for which the naïve analysis is also prone to bias. Investigators, reviewers, editors, and consumers of the literature all have a role to play in improving the quality of observational studies that rely on measures that are subject to systematic and random errors. More attention to these issues needs to be paid at the peer review stage. It is important that reviewers and editors be alert to the problems of measurement error and demand authors give some consideration of its impact in their research article. Further, as consumers of these studies, we need to take care to not cite or use the results of these studies without some acknowledgement of their limitations. It is perhaps only when an incomplete treatment of measurement error threatens the success of publication, that authors will be willing to invest the necessary effort into more fully addressing this limitation of their studies. With professional outreach, such as the activities of STRATOS that include preparation of a guidance paper on measurement error and misclassification and statistical methods to mitigate their bias, hopefully more investigators will have a better understanding of the impact of instrument bias and measurement error and the possible ways to address them.

Available data will determine which approaches of addressing measurement error are feasible for a given study. At a minimum, authors should state very explicitly the assumptions they are making about the structure of measurement error and the possible impact of those errors on their results. Using more formal analytical methods to adjust statistical analysis for measurement error generally require additional information, such as a reference instrument (unbiased measure of exposure on a subset). In some cases, repeat measurements are available on at least a subset of subjects, from which authors could consider to adjust analyses for within-person variability. If not available, sensitivity analyses, or bias analyses, can generally still be conducted to explore the potential impact of measurement error on study results and main conclusions.[7-11] Oftentimes there is at least some information about measurement properties of an instrument, such as from previous validation or reliability study, that can be used as a starting point for sensitivity analyses. In short, the current practice for presentation of results from studies with appreciable measurement error in the principle exposure measurement(s) needs to improve and in many studies this could be achieved using readily available resources and methods.

**Table 1** Literature search terms for the four topic areas: dietary intake cohort studies, dietary intake population survey, physical activity cohort study, and air pollution cohort studies.

|  | Search Terms |
|---|---|
| **Dietary Intake Cohort** | |
| Search A | ((((((((((((( (((cardiovascular disease[Title]) OR cancer[Title]) OR diabetes[Title])) AND (((risk[Title]) OR risks[Title]) OR association[Title])) AND ((((diet[Title]) OR consumption[Title]) OR intake[Title]) OR dietary[Title])) AND ("2012/01/01"[Date - Publication] : "3000"[Date - Publication]))) NOT case-control[Title/Abstract]) NOT review[Title/Abstract]) NOT meta-analysis[Title/Abstract])) NOT cross-sectional[Title/Abstract])) AND cohort)) |
| Search B1 | [Search A terms with date range extended to 2001/01/01] AND measurement error |
| Search B2 | (measurement error OR misclassification) AND nutritional epidemiology |
| Search B3 | (((((((measurement error[Title/Abstract] OR misclassification[Title/Abstract] OR reliability[Title/Abstract])) AND ((cardiovascular disease OR cancer OR diabetes))) AND ((risk OR association OR risks))) AND (diet[Title/Abstract] OR dietary[Title/Abstract] OR consumption[Title/Abstract] OR intake[Title/Abstract] OR intakes[Title/Abstract])) AND (cohort[Title/Abstract] OR men[Title/Abstract] OR women[Title/Abstract])) NOT meta-analysis) NOT case-control) NOT review |
| **Dietary Intake Population Survey** | |
| Search A | survey AND (nutrition OR nutritional OR diet OR dietary OR food OR nutrient) AND (FFQ OR FPQ OR record OR recall OR diary OR semi-quantitative OR semiquantitative). |
| Search B | N/A |
| **Physical Activity Cohort** | |
| Search A | survey AND (nutrition OR nutritional OR diet OR dietary OR food OR nutrient) AND (FFQ OR FPQ OR record OR recall OR diary OR semi-quantitative OR semiquantitative) |
| Search A | ((((("Cohort Studies"[Mesh:noexp]) OR "Follow-Up Studies"[Mesh]) OR "Longitudinal Studies"[Mesh]) OR "Prospective Studies"[Mesh] OR cohort OR prospective study)) AND (exercise OR recreation OR physical activity OR sedentary OR energy expenditure)) |
| Search B | [Search A terms] AND ("measurement error" OR misreport* OR misclassif* OR bias OR attenuat* OR calibrat*) |
| **Air Pollution Cohort** | |
| Search A | (("cardiovascular disease" OR cancer OR mortality OR "hospital admissions" OR "respiratory disease" OR diabetes OR biomarker OR  physiol* OR myocard* ) AND ( risk* OR association* OR effect* OR impact ) AND ( "air pollution" OR particulate* OR "environmental quality" OR "air quality" )) NOT ( case-control OR review OR meta-analysis OR cross-sectional ). |
| Search B1 | ((health OR "cardiovascular disease" OR cancer OR mortality OR "hospital admissions" OR "respiratory disease" OR diabetes OR biomarker OR  physiol* OR myocard* ) AND ( risk* OR association* OR effect* OR impact ) AND ( "air pollution" OR particulate* OR "environmental quality" OR "air quality" )) NOT ( case-control OR review OR meta-analysis OR cross-sectional ) AND ("measurement error" OR misreport* OR misclassif* OR bias OR attenuat* OR calibrat*) |
| Search B2 | "measurement error" AND "air pollution" |

**Table 2.** Study summary from general literature search without measurement error added to search terms (search A). Some questions were not assessed by all surveys.

|  | Dietary Intake Cohort N=51 | Physical Activity Cohort N=30 | Dietary Intake Population Survey N=67 | Air Pollution Cohort N=50 |
|---|---|---|---|---|
| Mention ME as potential problem N (%) | 48 (94%) | 17 (57%) | 53/67 (79%) | 21 (42%) |
| Included reliability substudy | 1 | 0 (0%) |  | 1 (2%) |
| Included calibration substudy | 2 | 0 (0%) |  | 1 (2%) |
| Used a method to adjust for ME N (%) | 5 (10%) | 0 (0%) | 19/67 (28%) | 3 (6%) |
| Categorized exposure[a] | Any 50/51 (98%) Exclusively 27/51 (53%) | Primary Exposure 21/30 (70%) |  |  |
| Statistic(s) of main interest[b] N (%) | HR 46 (90%) OR 3 (6%) RR 2 (4%) Slope 4 (8%) | HR 11 (37%) OR/RR 9 (30%) GLM 5 (17%) Other 5 (17%) | Mean 51 (76%) Median 28 (42%) %-tiles 21 (31%) Quality 31 (46%) |  |

[a]Articles were categorized as to whether they had categorized at least one dietary intake exposure in the statistical analysis (Any) and whether all analyses were done with categorized intakes (Exclusively).
[b]GLM: generalized linear model; HR: hazard ratio; ME: measurement error; OR: odds ratio; RR: relative risk.

**Table 3** Methods to address measurement error amongst articles. Results are from the methods review (search B) for the cohort surveys and the general search (search A) for the dietary intake survey.[a]

| Dietary Intake Cohort N= 27[b] | Physical Activity Cohort N=40 | Dietary Intake Population Survey N=67 | Air Pollution Cohort N = 25 |
|---|---|---|---|
| Any method 27 (100%) | Any method 2 (5%) | Any method 19 (28%) | Any method 5 (20%) |
| Reg Cal 26 (96%) | Reg Cal 2[c] (100%) | NCI  10 (53%) | Sens Anal 4 (80%) |
| SIMEX  1 (4%) | | Means 7 (37%) | Instr Var 1 (20%) |
| Other  1 (4%) | | ISU     1 (5%) | |
| | | MSM   1 (5%) | |
| | Search A: None 95% | | |
| Search A: None 90% | | Search A: None 72% | Search A: None 94% |

[a]Instr Var: Instrumental variables; ISU: Iowa State University Method[25]; MSM: Multiple Source Method[29,34]; NCI: National Cancer Institute Method[27,35]; Reg Cal: Regression calibration[1,22]; Sens Anal: Sensitivity analysis; SIMEX: Simulation Extrapolation Method[1,23].
[b]One article used both SIMEX and regression calibration so percentages do not add up to 100.
[c]One study did not use the term regression calibration but applied an equivalent method (i.e., beta coefficient adjustment for the intraclass correlation coefficient.[36]

**Table 4** Characteristics of articles reviewed for the Air Pollution Search A Survey (N=50).

| Main outcome | N (%) | Temporal resolution | N (%) | Type of measurement | N (%) |
|---|---|---|---|---|---|
| Mortality | 15 (30%) | Minutely | 1 (2%) | Fixed-site | 35 (70%) |
| Hospital admissions | 12 (24%) | Hourly | 9 (18%) | Personal | 5 (10%) |
| Cardiovascular disease | 2 (4%) | Between daily and hourly | 3 (6%) | Estimated exposure | 12 (24%) |
| Cancer | 1 (2%) | Daily | 21 (42%) | | |
| Respiratory disease | 7 (14%) | Weekly | 1 (2%) | | |
| Diabetes | 1 (2%) | Yearly | 3 (6%) | | |
| Physiological parameters | 5 (10%) | Study period | 4 (8%) | | |
| | | Other | 7 (14%) | | |
| Biomarker | 2 (4%) | | | | |
| Others | 6 (12%) | | | | |

**References**

1. Carroll RJ, Ruppert D, Stefanski LA, Crainiceanu CM. *Measurement Error in Nonlinear Models: A Modern Perspective*. Boca Raton: CRC press; 2006.

2. Gustafson P. *Measurement Error and Misclassification in Statistics and Epidemiology: Impacts and Bayesian Adjustments*. Boca Raton: CRC Press; 2003.

3. Fuller WA. *Measurement Error Models*. New York: John Wiley & Sons; 2009.

4. Buonaccorsi JP. *Measurement Error: Models, Methods, and Applications*. Boca Raton: CRC Press; 2010.

5. White E, Armstrong BK, Saracci R. *Principles of Exposure Measurement in Epidemiology: Collecting, Evaluating and Improving Measures of Disease Risk Factors*. Oxford: University Press; 2008.

6. Willett W. *Nutritional Epidemiology*. Oxford: Oxford University Press; 2012.

7. Fox MP. Creating a demand for bias analysis in epidemiological research. *J Epidemiol Community Health*. 2009 Feb 1;63(2):91.

8. Lash TL, Fox MP, MacLehose RF, Maldonado G, McCandless LC, Greenland S. Good practices for quantitative bias analysis. *Int J of Epidemiol*. 2014 Dec;43(6):1969-85.

9. Greenland S. Basic methods for sensitivity analysis of biases. *Int J Epidemiol*. 1996 Dec 1;25(6):1107-16.

10. Fox MP, Lash TL. On the Need for Quantitative Bias Analysis in the Peer-Review Process. *Am J Epidemiol*. 2017 May 15;185(10):865-8.

11. MacLehose RF, Olshan AF, Herring AH, Honein MA, Shaw GM, Romitti PA. Bayesian methods for correcting misclassification: an example from birth defects epidemiology. *Epidemiology*. 2009 Jan;20(1):27-35.

12. Freedman LS, Midthune D, Carroll RJ, Kipnis V. A comparison of regression calibration, moment reconstruction and imputation for adjusting for covariate measurement error in regression. *Stat Med*. 2008 Nov 10;27(25):5195-216.

13. Messer K, Natarajan L. Maximum likelihood, multiple imputation and regression calibration for measurement error adjustment. *Stat Med*. 2008 Dec 30;27(30):6332-50.

14. Cole SR, Chu H, Greenland S. Multiple-imputation for measurement-error correction. *Int J Epidemiol.* 2006 May 18;35(4):1074-81.

15. White IR. Commentary: Dealing with measurement error: multiple imputation or regression calibration? *Int J Epidemiol*. 2006 Jul 18;35(4):1081-2.

16. Bang H, Chiu YL, Kaufman JS, Patel MD, Heiss G, Rose KM. Bias Correction Methods for Misclassified Covariates in the Cox Model: comparison of five correction methods by simulation and data analysis. *J Stat Theory Pract*. 2013 Jan 1;7(2):381-400.

17. Jurek AM, Maldonado G, Greenland S, Church TR. Exposure-measurement error is frequently ignored when interpreting epidemiologic study results. *Eur J Epidemiol*. 2006 Dec 1;21(12):871-6.

18. https://epi.grants.cancer.gov/diet/usualintakes/macros.html, accessed on August 16, 2017.

19. https://epi.grants.cancer.gov/events/measurement-error/, accessed on August 16, 2017.

20. Sauerbrei W, Abrahamowicz M, Altman DG, Cessie S, Carpenter J. Strengthening analytical thinking for observational studies: The STRATOS initiative. *Stat Med*. 2014 Dec 30;33(30):5413-32.

21. Moher D, Liberati A, Tetzlaff J, Altman DG, The PRISMA Group (2009). Preferred Reporting Items for Systematic Reviews and Meta-Analyses: The PRISMA Statement. *J Clin Epidemiol*. 2009 Oct; 62 (10): 1006–1012.

22. Prentice RL. Covariate measurement errors and parameter estimation in a failure time regression model. *Biometrika*. 1982 Aug 1;69(2):331-42.

23. Cook JR, Stefanski LA. Simulation-extrapolation estimation in parametric measurement error models. J Am Stat Assoc. 1994 Dec 1;89(428):1314-28.

24. Beydoun MA, Kaufman JS, Sloane PD, Heiss G, Ibrahim J. n-3 Fatty acids, hypertension and risk of cognitive decline among older adults in the Atherosclerosis Risk in Communities (ARIC) study. *Public Health Nutr*.2008 Jan;11(1):17-29.

25. Nusser, SM, Carriquiry, AL, Dodd, KW and Fuller, WA. A semiparametric transformation approach to estimating usual daily intake distributions. *J Am Stat Assoc.* 1996; 91:1440-1449.

26. Nusser SM, Fuller WA, Guenther PM. Estimation of usual dietary intake distributions: Adjusting for measurement error and non-normality in 24-hour food intake data. In: Trewin D, ed. *Survey Measurement and Process Quality*. New York: Wiley; 1996: 689-709.

27. Tooze et al; A New Statistical Method for Estimating the Usual Intake of Episodically Consumed Foods with Application to Their Distribution. *J Am Diet Assoc*. 2006; 106:1575-1587.

28. Dekkers et al; SPADE, a New Statistical Program to Estimate Habitual Dietary Intake from Multiple Food Sources and Dietary Supplements. *J Nutr*. 2014; 144:2083-2091.

29. Haubrock J, Nöthlings U, Volatier JL, Dekkers A, Ocké M, Harttig U et al. (2011). Estimating usual food intake distributions by using the Multiple Source Method in the EPIC-Potsdam Calibration Study. *J Nutr* 2011; 141:914–920.

30. Zhang et al; A New Multivariate Measurement Error Model with Zero-Inflated Dietary Data, and its Application to Dietary Assessment. *Ann Appl Stat*. 2011; 5:1456–1487.

31. Mansournia MA, Danaei G, Forouzanfar MH, Mahmoodi M, Jamali M, Mansournia N, Mohammad K. Effect of physical activity on functional performance and knee pain in patients with osteoarthritis: analysis with marginal structural models. *Epidemiology*. 2012 Jul 1;23(4):631-40.

32. Gustafson P, Le ND. Comparing the effects of continuous and discrete covariate mismeasurement, with emphasis on the dichotomization of mismeasured predictors. *Biometrics*. 2002 Dec 1;58(4):878-87.

33. Freedman LS, Schatzkin A, Midthune D, Kipnis V. Dealing with dietary measurement error in nutritional cohort studies. *J Nat Can Inst.* 2011; 103:1086-1092.

34. Harttig U, Haubrock J, Knüppel S, Boeing H. 2011 The MSM program: web-based statistics package for estimating usual dietary intake using the Multiple Source Method. *Eur J Clin Nutr*. 65 S1:S87-9.

35. Tooze JA, Kipnis V, Buckman DW, Carroll RJ, Freedman LS, Guenther PM, Krebs-Smith SM, Subar AF, Dodd KW. A mixed-effects model approach for estimating the distribution of usual intake of nutrients: the NCI method. *Stat Med* 2010 Nov 30;29(27):2857-68.

36. Wientzek A, Tormo Díaz MJ, Castaño JM, Amiano P, Arriola L, Overvad K, Østergaard JN, Charles MA, Fagherazzi G, Palli D, Bendinelli B. Cross-sectional associations of objectively measured physical activity, cardiorespiratory fitness and anthropometry in European adults. *Obesity*. 2014 May 1;22(5).

**Supplemental Web Materials for "Epidemiologic analyses with error-prone exposures: Review of current practice and recommendations"**

-PRISMA Flow Diagrams for literature surveys

-Protocol document

# eFigure 1. PRISMA Flow Diagram: Dietary Intake Cohort Search A

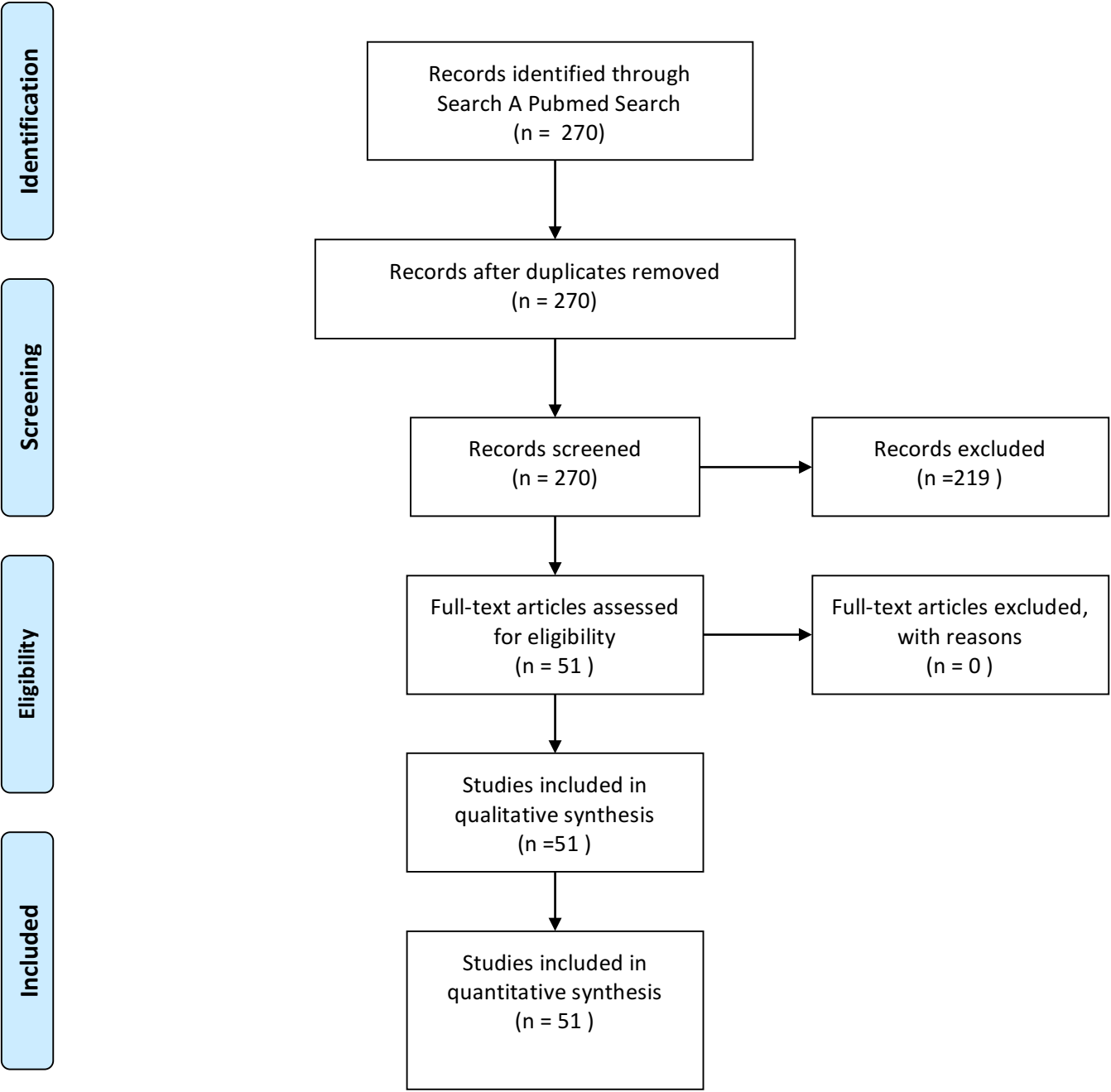

# eFigure 2. PRISMA Flow Diagram: Dietary Intake Search B

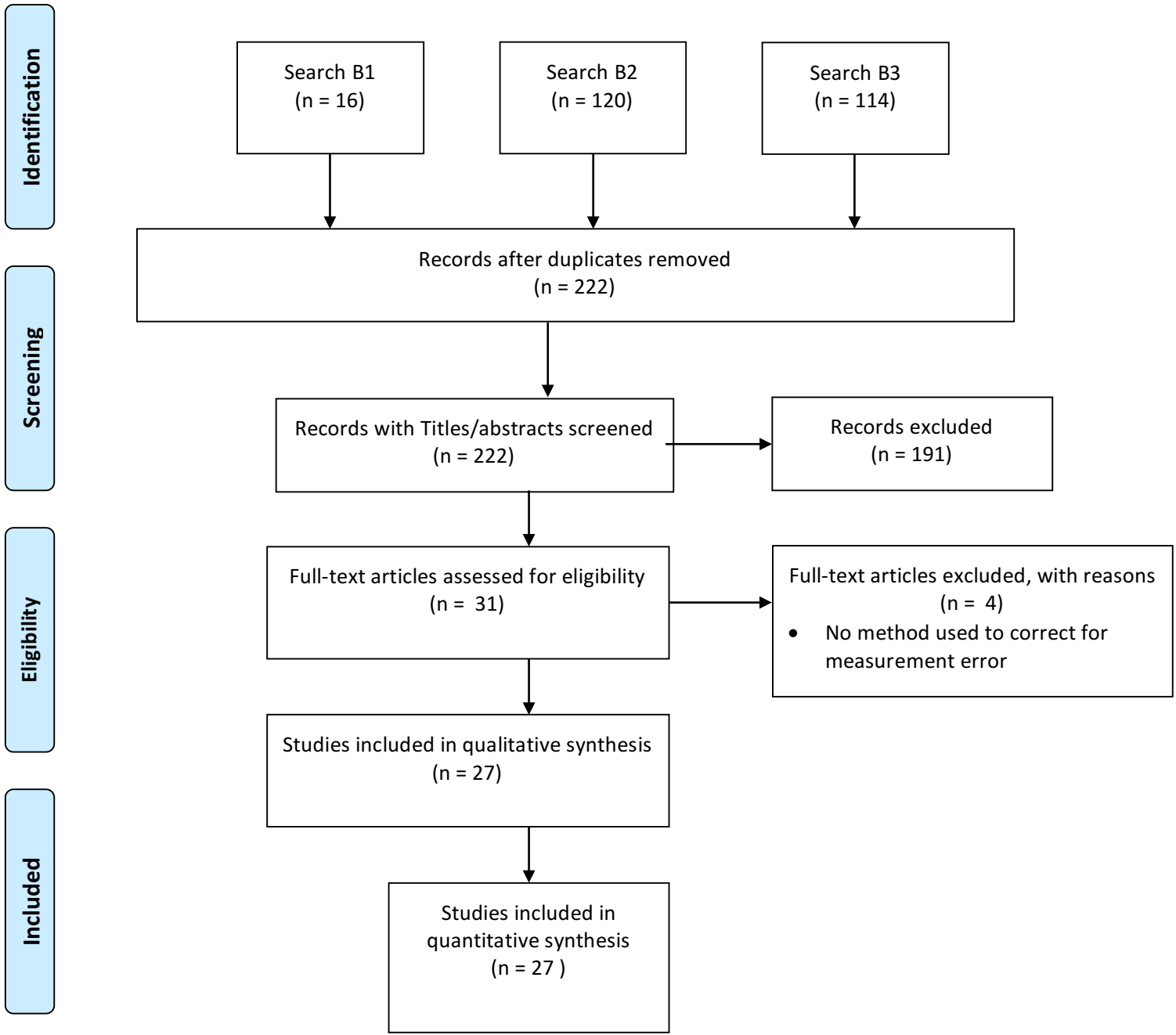

# eFigure 3. PRISMA Flow Diagram: Dietary Intake Survey Search A

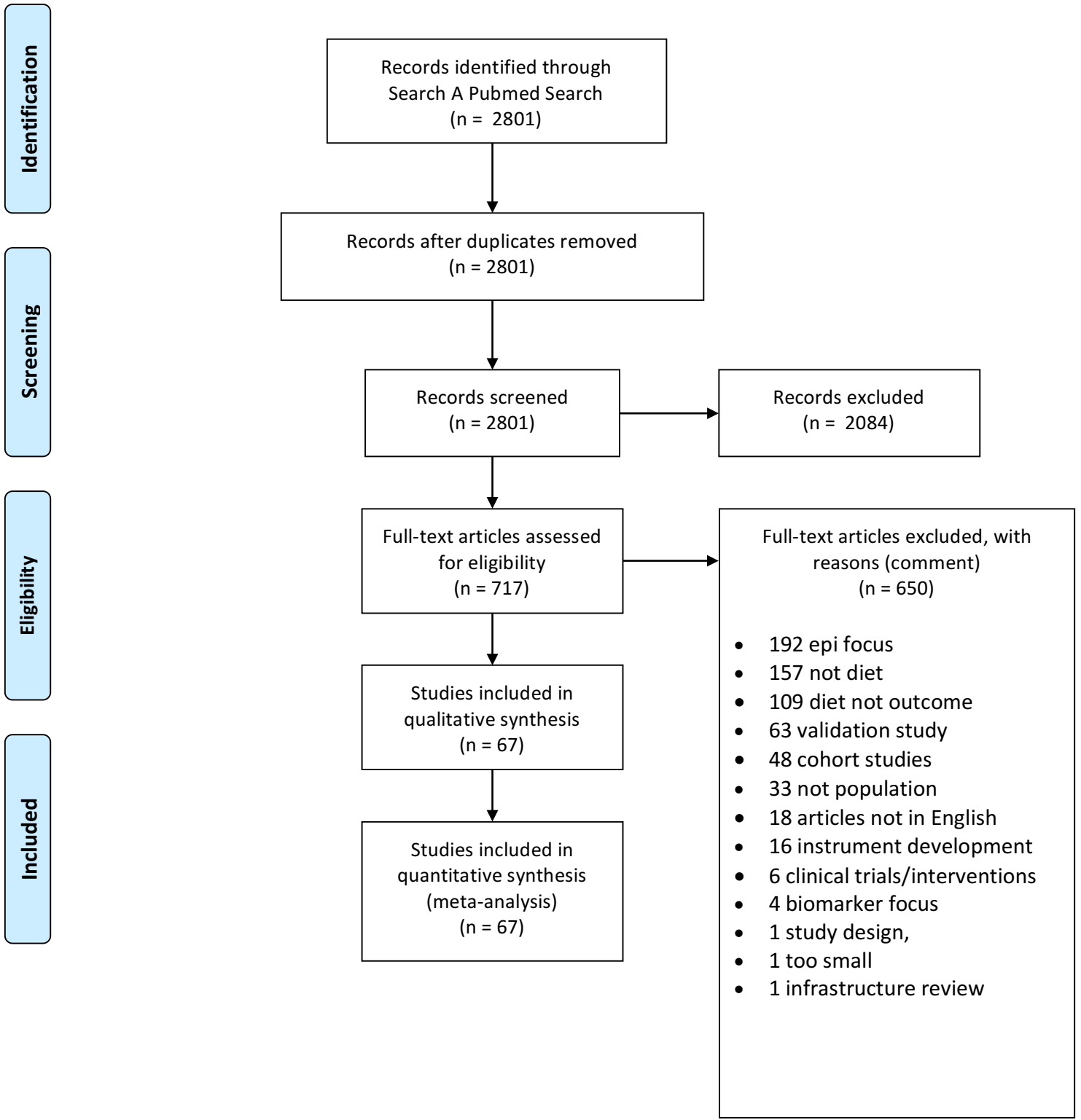

# eFigure 4. PRISMA Flow Diagram: Physical Activity Cohort Search A

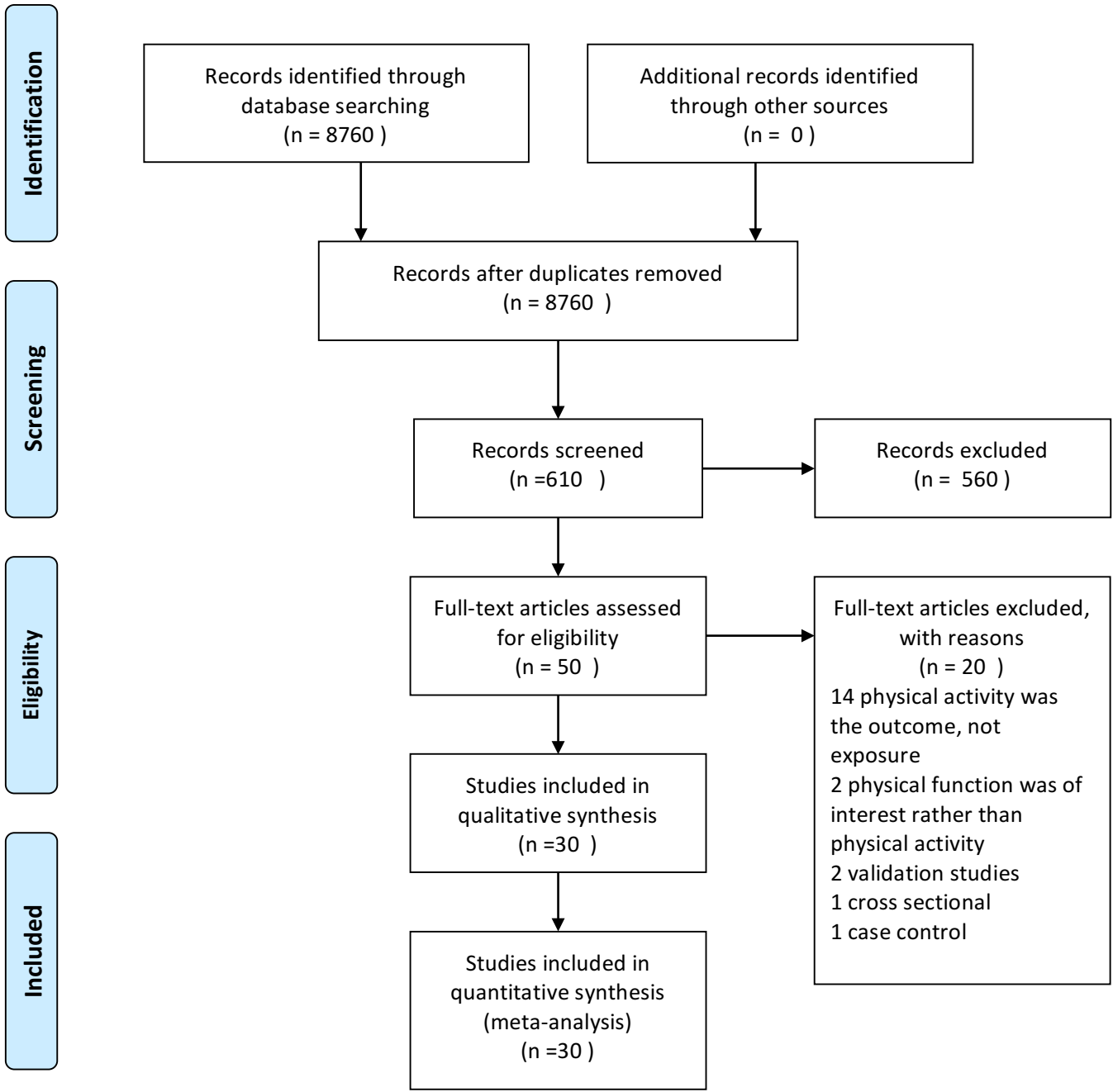

**eFigure 5. PRISMA Flow Diagram: Physical Activity Cohort Search B**

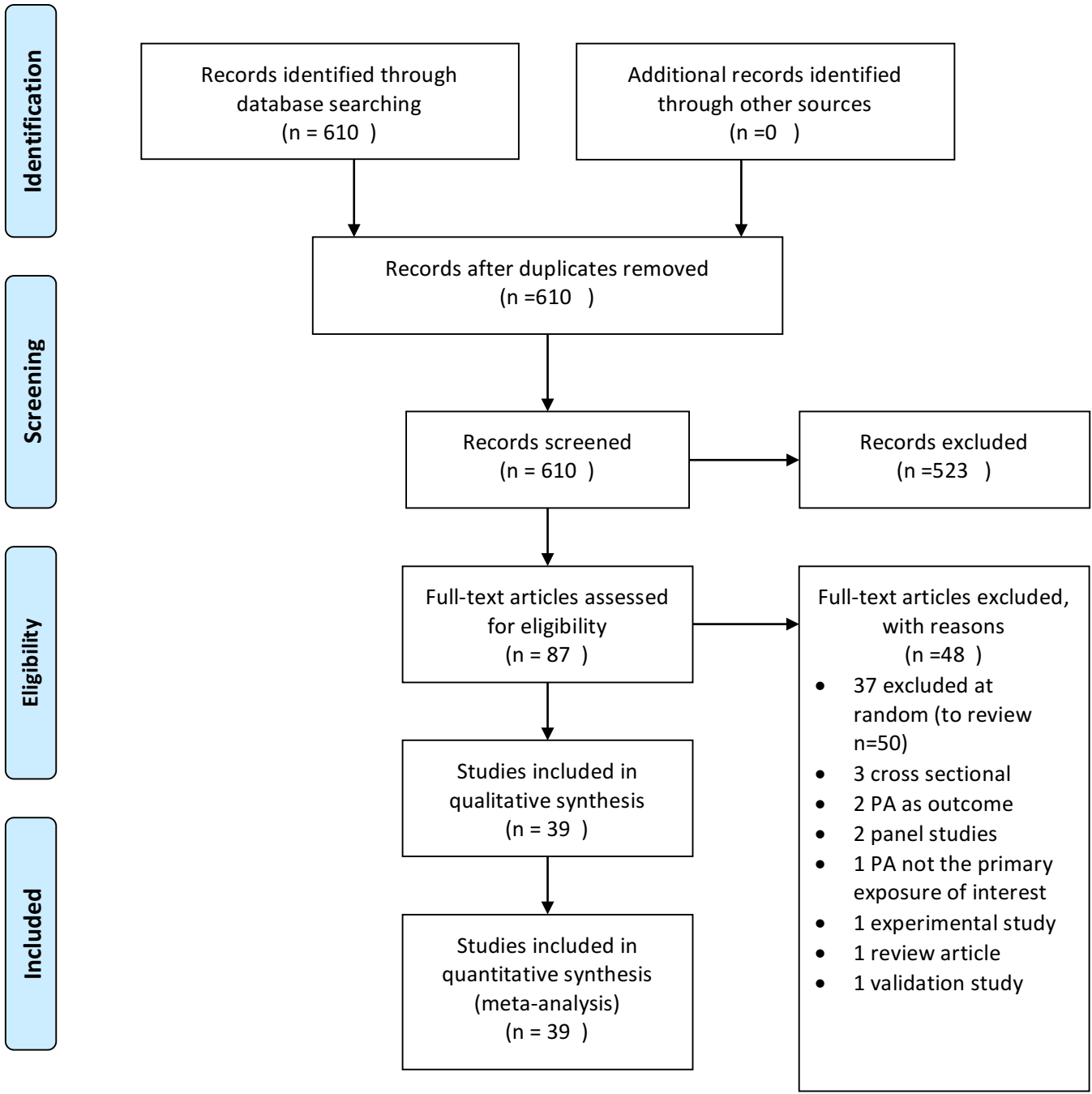

## eFigure 6. PRISMA Flow Diagram: Air Pollution Cohort Search A

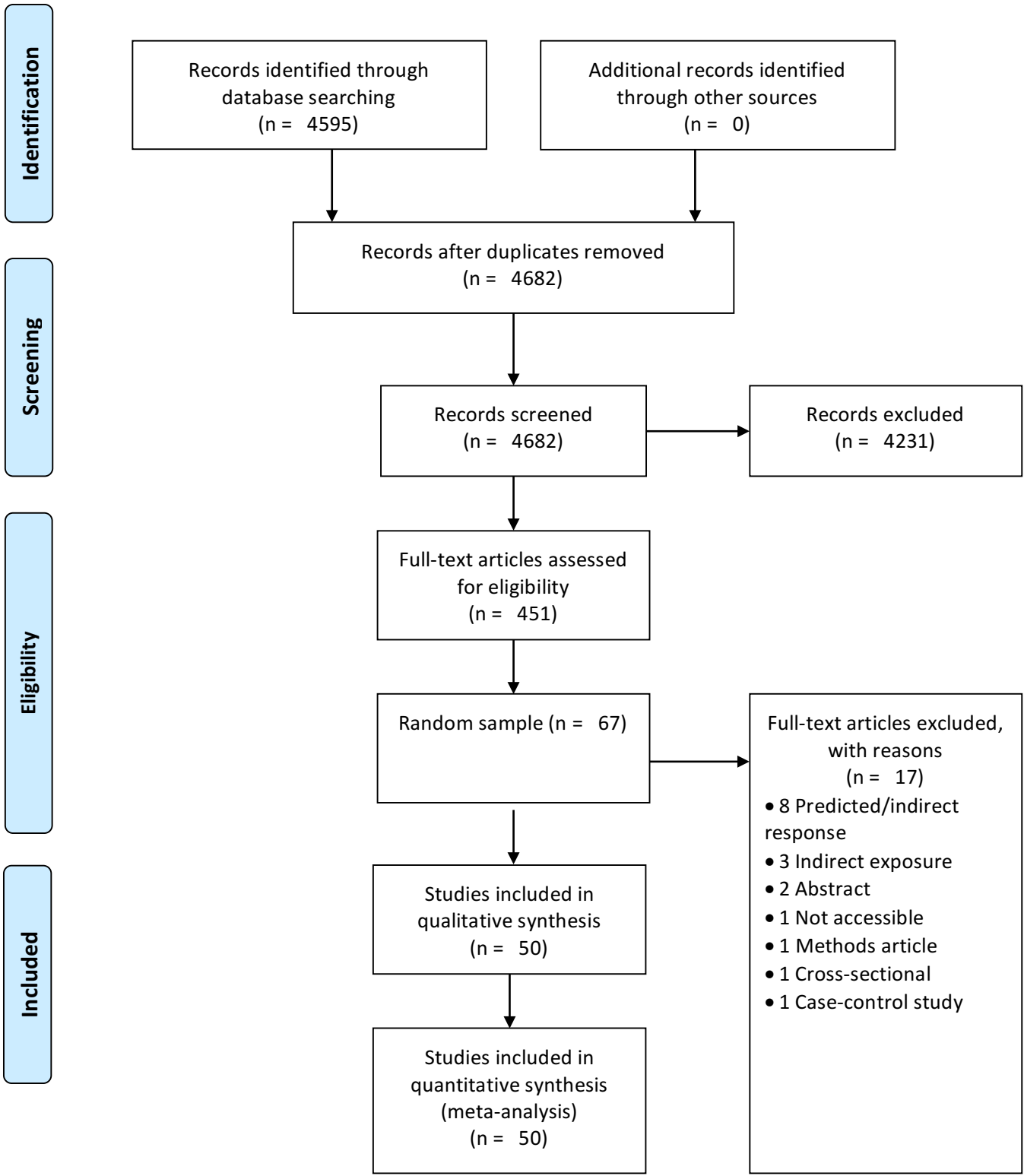

## eFigure 7. PRISMA Flow Diagram: Air Pollution Cohort Search B

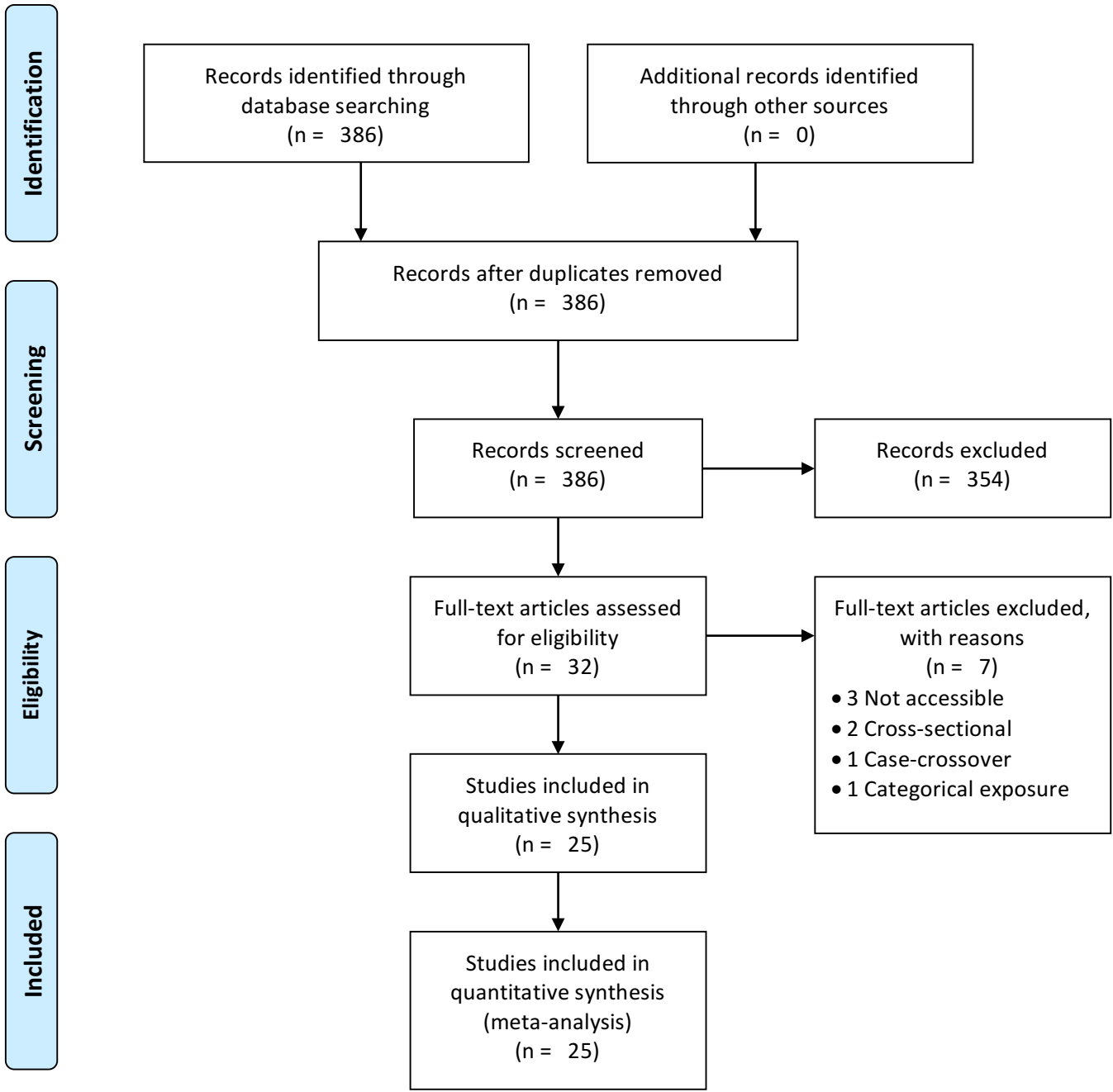

**STRATOS: Outline protocol for the nutrition and physical activity survey**

(a) Main questions:
(i) Nutritional epidemiology cohort studies:
   A. What proportion of reports mentioned dietary measurement error as a potential problem?
   B. What information was reported (relative risk for continuous intake or categorized intake)?
   C. What proportion included a calibration or repeated measurements sub-study (each type recorded separately) to allow for measurement error adjustment?
   D. What is the distribution of the size of the calibration and cohort studies?
   E. What type of reference instruments were used in such calibration studies, i.e., proportion using objective monitoring, subjective monitoring, or both.
   F. Was there adjustment for physical activity self-report data in the health outcome models? And if so, was the measurement error component of self-reported activity addressed in the model?
   G. What proportion mentioned accounting for measurement error in power calculations?
   H. What proportion of reports used a method to adjust risk estimates for measurement error?
   I. What statistical methods were used for such adjustment?
   J. Where were the results of such adjustment reported (in the abstract; in the main results section; in the discussion; in an appendix)?
   K. Was a statistician included as an author? or, if not, acknowledged for their help?

(ii) Population surveys of dietary intakes:
   A. What proportion or reports mentioned dietary measurement error as a potential problem?
   B. What information was reported (Means, Standard deviations, Percentiles)?
   C. What proportion included a calibration or repeated measurements sub-study (each type recorded separately), to allow for measurement error adjustment?
   D. What was the distribution of size of the calibration/repeat measurement sub-study?
   E. What reference instruments were used, i.e., proportion using objective monitoring, subjective monitoring, or both.
   F. What proportion of reports used a method to adjust the population distribution of intake for measurement error?
   G. What statistical methods were used for such adjustment?
   H. Where were the results of such adjustment reported (in the abstract; in the main results section; in the discussion; in an appendix)?
   I. Was a statistician included as an author? or, if not, acknowledged for their help?

(iii) Physical activity cohort studies:

A. What proportion of reports mentioned measurement error as a potential problem?
B. What information was reported (relative risk for continuous measure or categorized activity)?
C. What proportion included a calibration or repeated measurements sub-study (each type recorded separately) to allow for measurement error adjustment?
D. What is the distribution of the size of the calibration and cohort studies?
E. What type of reference instruments were used in the calibration studies, i.e proportion using objective monitoring, subjective monitoring, or both.
F. Was there adjustment for nutritional self-report data in the health outcome models? And if so, was the measurement error component of self-reported diet addressed in the model?
G. What proportion mentioned accounting for measurement error in power calculations or other aspects of study design?
H. What proportion of reports used a method to adjust risk estimates for measurement error?
I. What statistical methods were used for such adjustment?
J. Where were the results of such adjustment reported (in the abstract; in the main results section; in the discussion; in an appendix)?
K. Was a statistician included as an author? or if not, acknowledged for their help? If a statistician was not involved, what type of training did the person appear to have that did the statistical analysis (if possible to assess)?

L. What are the frequencies of the primary measures of interest (energy expenditure, PAL, AEE, minutes of activity, sedentary time, adherence to guidelines)?
M. What are the frequencies of the primary outcome of the study (diabetes, cancer, obesity, mortality, etc.)?
N. What types of models are used for the primary outcome (Cox regression, logistic regression, etc.)?
O. What study designs were found (prospective cohort, nested case control, case cohort)?
P. Were predictors treated as continuous or categorized in models?
Q. What proportion reported a statistically significant main analysis?
R. For physical activity calibration studies, what proportion of reports reported reliability, validity or coefficient of variation of their measure?

(b) Sampling Frame
Identify reports for each part of the survey by a literature search as follows:
A. Journals: Any
B. Years: 2012-14
C. Search engine: Pubmed or Scopus or maybe other; to be decided by survey leader.

- D. Keywords: To be developed by survey leader. Two sets of keywords: for surveying current practice, a set of keywords with no measurement error or misclassification terms; for surveying methods used, a set of keywords with measurement error and misclassification terms.
- E. Sample size aim: Pilot study phase: 10 papers for current practice and 10 papers for methods. Final sample size to be decided after pilot phase.
- F. Identification of "duplicate" reports.

(c) Data Collection:
- (i) There will be one reviewer per topic, except for nutritional cohorts where there will be two. The reviewers will check the reports that have been selected to verify that they should be included.
- (ii) Data collection forms will be developed for each part of the survey by the reviewers:
- (iii) The quality control person will review a 20% random sample per topic.
- (iv) Development of database. Survey leaders will develop their own databases.

(d) Personnel:
- (i) Laurence Freedman, Coordinator and quality control for nutritional surveys and physical activity
- (ii) Victor Kipnis, Co-Coordinator and quality control for nutritional cohorts
- (iii) Kevin Dodd, survey leader and reviewer for nutritional surveys
- (iv) Janet Tooze, survey leader and reviewer for physical activity
- (v) Pamela Shaw, survey leader and reviewer for nutritional cohorts
- (vi) Ruth Keogh, reviewer for nutritional cohorts
- (vii) Veronika Deffner, survey leader for air pollution
- (viii) Helmut Küchenhoff, reviewer for air pollution